\documentclass[lettersize,journal]{IEEEtran}
\usepackage{amsmath,amsfonts}
\usepackage{algorithmic}
\usepackage{algorithm}
\usepackage{array}
\usepackage[caption=false,font=normalsize,labelfont=sf,textfont=sf]{subfig}
\usepackage{textcomp}

\usepackage{stfloats}
\usepackage{url}
\usepackage{verbatim}
\usepackage{graphicx}
\usepackage{cite}
\usepackage{xcolor}
\usepackage{subfig}
\usepackage{titlesec}
\usepackage{mathrsfs}
\RequirePackage{CJKnumb}
\usepackage{amsmath}
\usepackage{amssymb}  
\usepackage{amsmath,amsfonts}
\usepackage{mathrsfs}
\usepackage{bm}

\newcommand{\bw}{\boldsymbol{w}}

\newcommand{\bd}{\boldsymbol{d}}

\begin{document}

\title{Secure Transmission for Fluid Antenna-Aided ISAC Systems}

\author{Yunxiao Li$^{\dagger}$, Qian Zhang$^{\dagger}$, \IEEEmembership{Graduate Student Member,~IEEE}, Xuejun Cheng,  Zhiguo Wang, \\ Xiaoyan Wang, Hongji Xu, \IEEEmembership{Member, IEEE}, Ju Liu, \IEEEmembership{Senior Member, IEEE}, 
\thanks{This work was supported by the National
	Natural Science Foundation of China under Grant 62071275. (Yunxiao Li and Qian Zhang contributed equally to this work.) ( \textit{Corresponding authors: Hongji Xu; Xiaoyan Wang}.) }
\thanks{Yunxiao Li , Qian Zhang, Xuejun Cheng, Zhiguo Wang, Hongji Xu, and Ju Liu are with the School of Information Science and Engineering, Shandong University, Qingdao 266237, China (e-mail: yunxiaoli@mail.sdu.edu.cn; qianzhang2021@mail.sdu.edu.cn; chengxuejun@mail.sdu.edu.cn; zhiguowang@mail.sdu.edu.cn; hongjixu@sdu.edu.cn; juliu@sdu.edu.cn). } 
\thanks{Xiaoyan Wang is with the Information Technology Service
	Center, People’s Court, Beijing 100745, China (e-mail: 428163395@139.com). }
\thanks{}
}

\markboth{}%
{Shell \MakeLowercase{\textit{et al.}}: A Sample Article Using IEEEtran.cls for IEEE Journals}

\IEEEpubid{}

\maketitle

\begin{abstract}
Fluid antenna (FA) has become a highly promising technology and has recently been used to enhance the integrated sensing and communication (ISAC) system. However, the scenario where sensing targets act as eavesdroppers in ISAC and how to maximize the sum secrecy rate has not been addressed. This letter investigates secure transmission in FA-aided ISAC systems, where the spatial agility of FAs enables enhanced physical layer security. We jointly optimize antenna position vector (APV) and beamforming to maximize the multiuser sum secrecy rate, which complicates the solution process. To solve the resulting non-convex problem, we use a block successive upper-bound minimization (BSUM) algorithm, which incorporates the proximal distance algorithm (PDA) for closed-form beamformer updates and extrapolated projected gradient (EPG) for APV optimization. Simulation results show that the proposed FA-ISAC scheme achieves over 20$\%$ sum secrecy rate gain compared to fixed-position antenna (FPA) systems.

\end{abstract}


\section{INTRODUCTION}
\IEEEPARstart{W}{ith}  the explosive growth in demand for Internet of Everything (IoE) applications, spectrum resources have become exceedingly scarce~\cite{9737357}. Integrated sensing and communication (ISAC) can effectively utilize signal waveforms to perform both communication and sensing functions, which has garnered significant attention to alleviate spectrum scarcity~\cite{8386661,10024901,9968163}. However, ISAC's dual-functional waveforms are vulnerable to eavesdropping, particularly
when sensing targets act as malicious eavesdroppers, necessitating robust security solutions.

In order to meet the ever-increasing security requirements, physical layer security (PLS), which is an important security technology, has received considerable attention, especially for ISAC systems~\cite{Zhang2021}. In~\cite{11165338}, Niu {\it et al.} provided a comprehensive survey on artificial noise techniques for PLS. In~\cite{9927490}, Yang {\it et al.} employed non-orthogonal multiple access to enable the ISAC network to accommodate more users, while leveraging precoding designs to improve security. However, existing studies mostly rely on fixed-position antennas (FPAs), which ignore spatial variability and limit secure transmission. Unlike FPAs, the spatial agility of fluid antennas (FAs) provides a flexible trade-off among communication rate, probing power, and information leakage when simultaneous sensing and secrecy are required ~\cite{11155198,10146274,11302793}. Notably, fluid antenna (FA) shares conceptual similarities with Flexible Intelligent Metasurfaces (FIM), as both exploit spatial reconfigurability and dynamic aperture manipulation to overcome static antenna limitations~\cite{zhang2026cram}. In~\cite{10751774}, the authors proposed a method that can reconstruct the full channel state information (CSI) corresponding to all FA position combinations from estimated CSI under the partial FA position combinations.
 In~\cite{10772590},  Zhang {\it et al.} considered a linear FA array-aided ISAC system, ensuring sensing performance while maximizing sum-rate through joint beamforming and antenna position optimization. In~\cite{ma2025movable}, Ma {\it et al.} proposed an MA-aided ISAC secure transmission system, where the security is improved by adjusting the antenna position at the user. Although existing research has demonstrated the effectiveness of FA in enhancing system security, the joint consideration of ISAC systems where the sensing target simultaneously acts as an eavesdropper, aimed at maximizing the multiuser sum secrecy rate via FA position optimization, remains largely unexplored.

This letter investigates the security of FA-aided ISAC systems, where the sensing target is considered as a potential eavesdropper. Our objective is to maximize the multiuser sum secrecy rate through the joint optimization of the beamformer and the positions of the FA. The optimization of beamforming vectors and antenna positions constitutes a non-convex problem, which complicates the solution process. To solve this problem, we use the block successive upper-bound minimization (BSUM) method to iteratively solve convex surrogate subproblems and employ the proximal distance algorithm (PDA) to derive a closed-form beamformer update in each iteration. Simultaneously, we derive the partial derivative of the eavesdropper rate with respect to the beamforming vector to obtain a closed-form solution for the optimization algorithm. Simulation results indicate that FAs can significantly  enhance the sum secrecy rate.

\section{SYSTEM MODEL}	
\index We consider an FA-aided ISAC downlink transmission system, as shown in Fig. 1. The base station (BS) equipped with $M$ FAs serves $K$ users. The sensing target acts as a potential eavesdropper targeting user private information, while interference is eliminated via isolation between the transmit and receive antennas. The BS transmission signal is
\begin{equation}\label{eqn-1} 
	\boldsymbol{x}_d= \sum\limits_{k=1}^{K}\boldsymbol{w}_k s_k,
\end{equation}
where $\boldsymbol{w}_k$ is the precoding vector for user $k$, and $s_k \sim {\cal CN}(0,1)$ represents the data symbol. In the ISAC system, the BS sends a dual-functional signal for simultaneous communication and sensing over block fading channels. We consider a linear FA port array of length $L$, within which the $m$-th FA can be reconfigured to position $d_m \in [0, L]$. Let $\boldsymbol{d} = [d_1, d_2, \ldots, d_M] \in \mathbb{R}^M$denote the antenna position vector (APV)\footnote{For simplicity, we consider a one-dimensional fluid antenna array. The proposed framework can be extended to 2-D fluid antenna arrays by appropriately reformulating the antenna position variables.}, which can be updated at a high refresh rate ~\cite{10146274}. The BS-user links are represented by a line-of-sight (LoS) model as
\begin{equation}\label{eqn-14} 
	\boldsymbol{h}_{k}= \bold a\left(\boldsymbol{d}, \phi_{k}\right)\delta_{k}, \quad k \in \mathcal{K} \triangleq\{1,2, ..., K\},
\end{equation}
where $\bold a (\boldsymbol{d}, \phi_{k}) = [e^{j v_{k} d_{1}}, e^{j v_{k} d_{2}}, \ldots, e^{j v_{k} d_{M}}]^{T}$, $v_{k}=\frac{2 \pi}{\lambda} \cos (\phi_{k})$.
 The wavelength is denoted by $\lambda$, the angle of departure (AOD) from the FA array toward user $k$ is represented by $\phi_{k}$, and the propagation gain is indicated by $\delta_k$.
Then, the $k$th user’s received signal is given by 
\begin{equation}\label{eqn-2} 
	\begin{split}
		y_{k}=\boldsymbol{h}_{k}^\mathrm{H} \boldsymbol{x}_{d}+n_{k}, \enspace \forall k \in \mathcal{K} .
	\end{split}
\end{equation}
The noise at user $k$ is represented by $n_{k} \sim \mathcal{C}\mathcal{N}(0, \sigma^2_k)$, and the signal-to-interference-plus-noise ratio (SINR) for user $k$ is defined accordingly,
\begin{equation}\label{eqn-2} 
	\gamma_{k}=\frac{\left|\boldsymbol{h}_{k}^{\mathrm{H}} \boldsymbol{w}_{k}\right|^{2}}{\sum_{i=1, i \neq k}^{K} \left|\boldsymbol{h}_{k}^{\mathrm{H}} \boldsymbol{w}_{i} \right|^{2} +      \sigma_{k}^{2}}, \quad k \in \mathcal{K}.
\end{equation} 

From the radar detection perspective, the BS-target channels $\boldsymbol{h}_{e}$ are also modeled by the following LoS model.	 
\begin{equation}\label{eqn-2} 
	\begin{split}
		\boldsymbol{h}_{e} = \delta_{e} \bold a(\boldsymbol{d},\phi) ,
	\end{split}
\end{equation}
where $\delta_{e}$ represents the propagation gain of the sensing target; $\bold a (\boldsymbol{d}, \phi) = [e^{j v d_{1}}, e^{j v d_{2}}, \ldots, e^{j v d_{M}}]^{T}$; $v=\frac{2 \pi}{\lambda} \cos (\phi)$ and $\phi$ is the AOD of the FA array along the probed direction. The SINR for user $k$ at the sensing target is 
\begin{equation}\label{eqn-2} 
	\gamma_{e, k}=\frac{\left|\boldsymbol{h}_{e}^{\rm H} \boldsymbol{w}_{k}\right|^{2}}{\sum_{j=1,j \neq k}^{K}\left|\boldsymbol{h}_{e}^{\rm H} \boldsymbol{w}_{j}\right|^{2}+\sigma_{e}^{2}},
\end{equation}
where $n_{e}\sim \mathcal{C}\mathcal{N}(0,\sigma^2_e)$ denotes the noise at the sensing target.

\begin{figure}[!t]
	\centering
	\includegraphics[width=2.3 in]{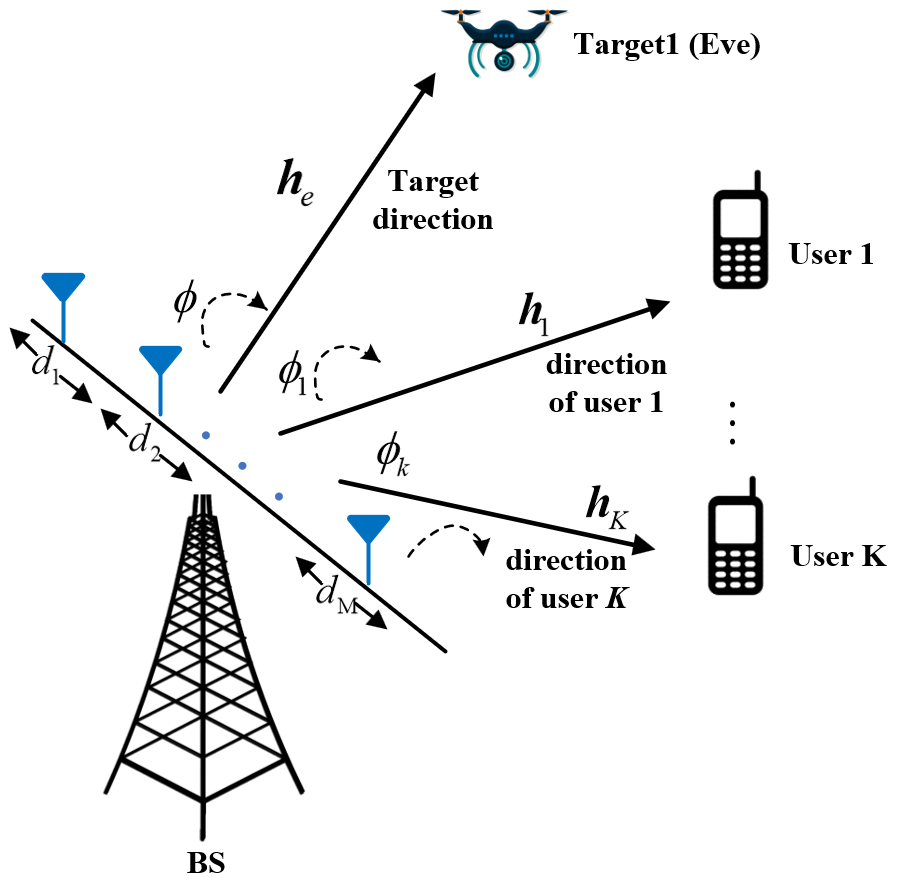}
	\caption{Secure transmission for FA-aided ISAC systems model.}
	\label{fig1}
	\vspace{-0.5cm}
\end{figure}
In the proposed ISAC framework, the transmit beamformer is additionally capable of sensing the existence or status of a target along the probing direction~\cite{8386661,4276989}. Consequently, the ISAC system must deliver adequate beam gain in the direction of interest. To characterize this, the transmission signal covariance matrix $\mathbf{R}_{w}$ is expressed as
\begin{equation}\label{eqn-2} 
	\mathbf{R}_{w}=\mathbb{E}\left[\boldsymbol{x}_{d} \boldsymbol{x}_{d}^{\mathrm{H}} \right]=\sum_{k = 1}^{K} \boldsymbol{w}_{k} \boldsymbol{w}_{k}^{\mathrm{H}} .
\end{equation}
We let the probing power along the desired direction be greater than a threshold.
\begin{align}
	P(\boldsymbol{w}, \boldsymbol{d}, \phi)=\mathrm{tr}( \bold a^{\rm H}(\boldsymbol{d}, \phi)  \mathbf{R}_{w} \bold a (\boldsymbol{d}, \phi)) \geq P_{d},
\end{align}
 where $\boldsymbol{w} = [\boldsymbol{w}_1^\mathrm{T},\ldots, \boldsymbol{w}_M^\mathrm{T} ]^{\rm \mathrm{T}}$. In addition, the sum secrecy rate corresponding to user $k$ can be expressed as
\begin{equation}\label{eqn-2} 
	C_{s}^{k}=\left[\log _{2}\left(1+\gamma_{k}\right)-\log _{2}\left(1+\gamma_{e, k}\right)\right]^+,
\end{equation}
where the $[x]^+$ denotes the positive part of $x$. 

\section{BEAMFORMER AND APV OPTIMIZATION}
Our objective is to jointly optimize the beamformer and APV to maximize the sum secrecy rate. The corresponding joint design problem is formulated as follows.
\begin{subequations}\label{eq:BCD}
	\begin{align}
		&\max_{\boldsymbol{w}, \boldsymbol{d}} \sum_{k = 1}^{K} C_{s}^{k} \tag{10a}\\
		\text{s.t.} &\quad \mathcal{C}_{\sf d}: P(\boldsymbol{w}, \boldsymbol{d}, \phi) \geq P_{d} \tag{10b}, \\	
		&\quad \mathcal{C}_{\sf FA}: d \geq 0,\quad d_{M} \leq L \tag{10c},\\
		&\quad \mathcal{C}_{\sf SP}: d_{m} - d_{m - 1} \geq L_{0}, \quad m = 2,3, \cdots, M \tag{10d},\\ 
		&\quad \mathcal{C}_{\sf BS}: \sum_{k = 1}^{K}\left\| \boldsymbol{w}_{k}\right\| _{2}^{2}  \leq P_{max} \tag{10e}.                                                                             
	\end{align}
\end{subequations}

 Constraint (10b)  imposes a beam gain threshold of the desired sensing direction. Constraints (10c) and (10d) constrain the antenna positions, guaranteeing they lie within the interval $[0, L]$ and that the distance between any two antennas is greater than the minimum antenna separation $L_0$. Constraint (10e) specifies the transmit power constraint. Observe that the optimization problem (10) exhibits significant non-convexity relative to the design variables ($\boldsymbol{w}$, $\boldsymbol{d}$). 
To develop an efficient algorithm for solving problem (10), we exploit the block variable structure and extend the BSUM method to address it. The primary difficulty involves integrating the FA positions into this process. We transform the maximization problem into an equivalent minimization problem. The key steps of the tailored algorithm are as follows~\cite{10446199,5756489}.
\begin{align}
	\min_{\boldsymbol{w},\boldsymbol{d},\boldsymbol{u},\boldsymbol{\xi},\boldsymbol{\rho},\boldsymbol{\eta}} 
	& \mathcal{F}_k(\boldsymbol{w}, \boldsymbol{d}, \boldsymbol{u}, \boldsymbol{\rho})+\mathcal{F}_e(\boldsymbol{w}, \boldsymbol{d}, \boldsymbol{\xi}, \boldsymbol{\eta}) \\
	\text{ s.t.} 
	& \quad \mathcal{C}_{\sf d}, ~\mathcal{C}_{\sf FA}, ~\mathcal{C}_{\sf SP}, ~\mathcal{C}_{\sf BS},
\end{align}
where $\mathcal{F}_k$ represents the communication rate at users and $\mathcal{F}_e$ represents the communication rate of the sensing target. 
\begin{equation*}
\begin{split}
	&\mathcal{F}_k(\boldsymbol{w}, \boldsymbol{t}, \boldsymbol{u}_k, \boldsymbol{\rho}) = \sum_{k=1}^{K}[ - 2\mathscr{R}\{\boldsymbol{b}_k^\mathrm{H}(\boldsymbol{d})\boldsymbol{w}_k\} - \log(\rho_k)] \\
	&\qquad\qquad\qquad\qquad\qquad\qquad +\sum_{k=1}^{K}[\boldsymbol{w}_k^\mathrm{H} \bold{A}(\boldsymbol{d})\boldsymbol{w}_k],\\
	&\mathcal{F}_e(\boldsymbol{w}, \boldsymbol{d}, \boldsymbol{u}_e, \boldsymbol{\rho}) =    \sum_{k=1}^{K} [-2\mathscr{R} \{\boldsymbol{g}_k^\mathrm{H}(\boldsymbol{d})\boldsymbol{w}_k\}  - \log(\rho_k)]  \\
	&\qquad\qquad\qquad\qquad\qquad\qquad   +\sum_{k=1}^{K} \sum_{i=1, i\neq k}^{K} [\boldsymbol{w}_i^\mathrm{H} \bold{E}_k(\boldsymbol{d}) \boldsymbol{w}_i],
	\end{split}
\end{equation*}
where $\bold{A}(\boldsymbol{d}) = \sum_{k=1}^{K} \rho_k|u_k|^2\boldsymbol{h}_k(\boldsymbol{d})\boldsymbol{h}_k^\mathrm{H}(\boldsymbol{d})$;  $\bold{E}_k(\boldsymbol{d}) =\eta_{k}|\xi_{k}|^2\boldsymbol{h}_{e}(\boldsymbol{d})\boldsymbol{h}_{e}^\mathrm{H}(\boldsymbol{d})$; $\boldsymbol{b}_k(\boldsymbol{d}) = \rho_k u_k \boldsymbol{h}_k(\boldsymbol{d})$; $\boldsymbol{g}_k(\boldsymbol{d}) =j \eta_k \xi_k \boldsymbol{h}_{e}(\boldsymbol{d})$. Auxiliary variables $\boldsymbol{u} \in \mathbb{C}^{K}$, $\boldsymbol{\xi} \in \mathbb{C}^{K}$, $\boldsymbol{\rho} \in \mathbb{R}^{K}_{++}$ and  $\boldsymbol{\eta} \in \mathbb{R}^{K}_{++}$ are introduced, and $\mathscr{R}\{x\}$ represents the real part of $x$. By utilizing this transformation, we employ the BSUM framework to derive the following
\begin{subequations}\label{eq:BCD}
	\begin{align}
	&\boldsymbol{w}^{\ell+1} =  \arg \min_{\boldsymbol{w} \in \mathcal{C}_{\sf d} \cap \mathcal{C}_{\sf BS}} \mathcal{F} (\boldsymbol{w}, \boldsymbol{d}^\ell), \\
	&\boldsymbol{d}^{\ell+1} = \arg \min_{\boldsymbol{d} \in \mathcal{C}_{\sf d} \cap ~\mathcal{C}_{\sf FA} \cap ~\mathcal{C}_{\sf SP}} \mathcal{F} (\boldsymbol{w}^{\ell+1}, \boldsymbol{d}),             
	\end{align}
\end{subequations}
where $\boldsymbol{u}$ , $\boldsymbol{\xi}$ , $\boldsymbol{\rho}$ and $\boldsymbol{\eta}$ are given as $u_k^{\ell+1} = \frac{\boldsymbol{h}_k^\mathrm{H} \boldsymbol{w}_k}{\sum_{i=1}^{K} |\boldsymbol{h}_k^\mathrm{H} \boldsymbol{w}_i|^2 + \sigma_k^2}$ , $\xi_{k}^{\ell+1} = \frac{j \boldsymbol{h}_{e}^\mathrm{H} \boldsymbol{w}_k}{\sum_{i=1,i\neq k}^{K} |\boldsymbol{h}_{e}^\mathrm{H} \boldsymbol{w}_i|^2 + \sigma_e^2}$ , $\rho_k^{\ell+1} = \frac{1}{1 - (u_k^{\ell+1})^* \boldsymbol{h}_k^\mathrm{H} \boldsymbol{w}_k}$ and $\eta_{k}^{\ell+1} = \frac{1}{1 - j (\xi_k^{\ell+1})^* \boldsymbol{h}_{e}^\mathrm{H} \boldsymbol{w}_k}$, for any $k \in \mathcal{K}$, where $(\cdot)^*$ represents the complex conjugate operation. These updates follow from the KKT first-order optimality conditions of the transformed objective.

We will provide efficient algorithms for optimizing the subproblems of $\boldsymbol{w}$ and $\boldsymbol{d}$ below.

\subsection {Beamformer Optimization}  
To address the subproblem in (13a), we introduce a PDA that effectively decouples the constraints and enables efficient closed-form updates. However, this projection gradient calculation is very complicated. Using the PDA algorithm, we first express problem (13a) as
\begin{equation}
	\min_{\boldsymbol{w}} \mathcal{F}_k (\boldsymbol{w}) + \mathcal{F}_e(\boldsymbol{w}) + \bar{\mu} \,    \text{dist}^2(\boldsymbol{w}, \mathcal{C}_{\sf d}) +\bar{\mu} \, \text{dist}^2(\boldsymbol{w}, \mathcal{C}_{\sf BS}),
\end{equation}
where $\text{dist}(\boldsymbol{w}, \mathcal{X})$ represents the operation of projecting point  $\boldsymbol{w}$ onto set $\mathcal{X}$, and $\bar{\mu} > 0$ is a prescribed penalty parameter. It is evident that, when $\bar{\mu}$ is sufficiently large, the optimal solution of problem (14) coincides with that of problem (13a).

Additionally, the distance function lacks a closed-form expression. Therefore, we address it by constructing a convex upper bound of the distance function,
\begin{align}
	\text{dist}(\boldsymbol{w}, \mathcal{C}_{\text{BS}}  \cap \mathcal{C}_{\text{d}}) \leq \text{dist}(\boldsymbol{w}, \mathcal{C}_{\text{BS}} )  + \text{dist}(\boldsymbol{w}, \mathcal{C}_{\text{d}}) ).
\end{align}
Perform a first-order upper bound on each distance function,
\begin{align}
	\text{dist}(\boldsymbol{w}, \mathcal{C}_{\text{BS}}) \leq  &\left\| \boldsymbol{w} - \tilde{\boldsymbol{w}}_{\text{BS}} \right\|_2^2, \\
	\text{dist}(\boldsymbol{w}, \mathcal{C}_{\text{d}}) \leq  &\left\| \boldsymbol{w} - \tilde{\boldsymbol{w}}_{\text{d}} \right\|_2^2,
\end{align}
where $\tilde{\boldsymbol{w}}_{\text{BS}} = \Pi_{{\cal C}_{\sf BS}}(\bw)$ and  $\tilde{\boldsymbol{w}}_{\text{d}} = \Pi_{{\cal C}_{d}}(\bw)$, while  $\Pi_{\mathcal{X}} (\boldsymbol{w}) = \mathrm{arg \   min}_{\boldsymbol{y} \in \mathcal{X} } \|\boldsymbol{x}-\boldsymbol{y}\|_2^{2}$ is the projection of $\boldsymbol{x}$ onto the set $\mathcal{X}$ and
\begin{equation*}
	\begin{aligned}
		[\widetilde{\boldsymbol{w}}_{\sf BS}]_k &= 
		\begin{cases}
			\boldsymbol{w}_k, & \text{if } \sum_{k=1}^{K} \|\boldsymbol{w}_k\|_2^2 \leq P_{max}, \\
			\sqrt{\frac{P_{max}}{\sum_{k=1}^K \|\boldsymbol{w}_k\|_2^2}}\boldsymbol{w}_k, & \text{otherwise},
		\end{cases}\\
		[\widetilde{\boldsymbol{w}}_{\sf d}]_k &= 
		\begin{cases}
			\boldsymbol{w}_k, \quad \quad \text{if } \sum_{k=1}^{K} \boldsymbol{w}_k^\mathrm{H} \boldsymbol{a}(\boldsymbol{d}, \phi)\boldsymbol{a}^\mathrm{H}(\boldsymbol{d}, \phi)\boldsymbol{w}_k \geq P_d, \\
			(\frac{1}{\boldsymbol{I}-  \mu\boldsymbol{a}(\boldsymbol{d}, \phi)\boldsymbol{a}^\mathrm{H}(\boldsymbol{d}, \phi)\boldsymbol{w}_k} , \quad\quad \text{otherwise},
		\end{cases}
	\end{aligned}	
\end{equation*}
where $\mu$ is the Lagrange multiplier and $\mu = \frac{1}{\|\boldsymbol{a}(\boldsymbol{d}, \phi)\|_2^2} - \sqrt{\frac{\sum_{k=1}^{K} \boldsymbol{w}_k^\mathrm{H} \boldsymbol{a}(\boldsymbol{d}, \phi)\boldsymbol{a}^\mathrm{H} (\boldsymbol{d}, \phi)\boldsymbol{w}_k} {|\boldsymbol{a}^\mathrm{H}(\boldsymbol{d}, \phi)\boldsymbol{a} (\boldsymbol{d}, \phi)|^2 P_d}}$~\cite{10446199}.

With the distance majorization, at iteration $\boldsymbol{w}$, the problem reduces to solving an unconstrained quadratic program formulated as
\begin{equation*}
	\begin{aligned}
		\min_{\boldsymbol{w}} \mathcal{F}_k (\boldsymbol{w}) + \mathcal{F}_e(\boldsymbol{w}) + \bar{\rho}\left(\|\boldsymbol{w} - \widetilde{\boldsymbol{w}}_{\text{BS}}\|_2^2 + \|\boldsymbol{w} - \widetilde{\boldsymbol{w}}_\text{d}\|_2^2\right),
	\end{aligned}	
\end{equation*}
which can be solved in closed-form to obtain the optimal solution, i.e.,
\begin{equation} \label{gradient_method}
	\begin{split}
		\boldsymbol{w}_k = (&\bold{A}(\boldsymbol{d})+\bold{E}(\boldsymbol{d}) - \bold{E}_i(\boldsymbol{d}) + 2\bar{\rho}I)^{-1}    \\ 
		&\left[\boldsymbol{b}_k(\boldsymbol{d}) + \boldsymbol{d}_e(\boldsymbol{d}) +  \bar{\rho}\left([\tilde{\boldsymbol{w}}_{\text{BS}}]_k + [\tilde{\boldsymbol{w}}_{\text{d}}]_k \right) \right]. 
	\end{split}
\end{equation}

The derivation of the projection $\boldsymbol{w}_k$ of the $\mathcal{F}_e(\boldsymbol{w})$ is shown in Appendix A.
\subsection {APV Optimization}  
We use an extrapolated projected gradient (EPG) algorithm to address problem (13b), which takes the following updates
\begin{equation} \label{gradient_method}
	\begin{split}
		&\tilde{\boldsymbol{d}}_{i+1} = \Pi_{\mathcal{C}_{\sf d} \cap \mathcal{C}_{\sf FA} \cap \mathcal{C}_{\sf SP}} \left( \boldsymbol{\lambda}_i - \alpha \nabla_d \mathcal{F}(\boldsymbol{\lambda}_i | \boldsymbol{d}_i) \right), \\
		& \boldsymbol{d}_{i+1} = \tilde{\boldsymbol{d}}_{i+1} + \eta_{i+1} \left( \tilde{\boldsymbol{d}}_{i+1} - \boldsymbol{d}_i \right),
	\end{split}
\end{equation}	
where $\alpha > 0$ is a step determined through backtracking line search. The gradient $\nabla_{\boldsymbol{d}} \mathcal{F}(\boldsymbol{\lambda}_i | \boldsymbol{d}_i)$ can be computed as presented in the following. The  extrapolation coefficient $\eta_{i+1}$ is defined as $\eta_{i+1} = \frac{\varsigma_{i+1} - 1}{\varsigma_{i+1}},$ and the parameter \(\varsigma_{i+1}\) is updated via as $\varsigma_{i+1} = \frac{1 + \sqrt{1 + 4 \varsigma_i^2}}{2} $, $ \varsigma_1 = 0.$
\begin{align}
	\nabla_{\boldsymbol{d}} \mathcal{F} &(\boldsymbol{d}) = \sum_{k=1}^{K}  \biggl\{ \rho_k|u_k|^2\delta_k^2 \sum_{i=1}^{K} \frac{\partial f_{k,i}}{\partial \boldsymbol{d}}  
	- 2\mathscr{R}\{\rho_ku_k^*\delta_k\}\frac{\partial h_{k}}{\partial \boldsymbol{d}} \biggr.  \notag \\
	&\biggl. +\eta_k|\xi_k|^2\delta_e^2 \sum_{i=1}^{K} \frac{\partial p_{k,i}}{\partial \boldsymbol{d}} - 2\mathscr{R}\{j \eta_k \xi_k^*\delta_e\}\frac{\partial l_{k}}{\partial \boldsymbol{d}}\biggr\}, 
\end{align}	
 where 
\begin{equation*}
	\begin{aligned}
		\frac{\partial f_{k,i}}{\partial \boldsymbol{d}} 
		& =2v_k [\text{diag}(\boldsymbol{g}_k)(\boldsymbol{C}_i\boldsymbol{q}_k  -  \boldsymbol{g}_k^T \boldsymbol{D}_i)
		\\& \qquad\qquad\qquad\qquad\qquad- \text{diag}(\boldsymbol{q}_k)(\boldsymbol{C}_i\boldsymbol{g}_k +  \boldsymbol{D}_i \boldsymbol{q}_k )],\\
		\frac{\partial h_{k}}{\partial \boldsymbol{d}} 
		&=v_k\left[ \mathrm{{diag}}(\boldsymbol{g}_k)\boldsymbol{b}_k - \mathrm{{diag}}(\boldsymbol{q}_k)\boldsymbol{a}_k  \right],\\
		\frac{\partial p_{k,i}}{\partial \boldsymbol{d}} 
		&=2v[\text{diag}(\boldsymbol{r})(\boldsymbol{C}_i\boldsymbol{u}  -  \boldsymbol{r}^T \boldsymbol{D}_i)-\text{diag}(\boldsymbol{u})(\boldsymbol{C}_i\boldsymbol{r} +  \boldsymbol{D}_i \boldsymbol{u} )],\\
		\frac{\partial l_{k}}{\partial \boldsymbol{d}} 
		&=v\left[ \mathrm{{diag}}(\boldsymbol{r}_k)\boldsymbol{b}_k - \mathrm{{diag}}(\boldsymbol{u}_k)\boldsymbol{a}_k  \right],\\
		f_{k,i} &= |\boldsymbol{a}^\mathrm{H}(\boldsymbol{d}, \phi_k)\boldsymbol{w}_i|^2  = \boldsymbol{g}_k^\mathrm{T} \bold{C}_i \boldsymbol{g}_k + \boldsymbol{q}_k^\mathrm{T} \bold{C}_i \boldsymbol{q}_k + 2\boldsymbol{g}_k^\mathrm{T} \bold{D}_i \boldsymbol{q}_k , \\
		h_{k} &= \mathscr{R}\{\boldsymbol{a}^\mathrm{H}(\boldsymbol{d}, \phi_k)\boldsymbol{w}_k\} = \boldsymbol{g}_k^\mathrm{T} \boldsymbol{a}_k  + \boldsymbol{q}_k^\mathrm{T} \boldsymbol{b}_k,\\
		p_{k,i} &= |\boldsymbol{a}^\mathrm{H}(\boldsymbol{d}, \phi)\boldsymbol{w}_i|^2 = \boldsymbol{r}^\mathrm{T}\bold{C}_i\boldsymbol{r}  +  \boldsymbol{u}^\mathrm{T} \bold{C}_i\boldsymbol{u}  + 2 \boldsymbol{r}^\mathrm{T}\bold{D}_i\boldsymbol{u},\\
		l_{k} &= \mathscr{R}\{\boldsymbol{a}^\mathrm{H}(\boldsymbol{d}, \phi)\boldsymbol{w}_k\} =  \boldsymbol{r}^\mathrm{T}\boldsymbol{a}_k + \boldsymbol{u}^\mathrm{T}\boldsymbol{b}_k.
	\end{aligned}
\end{equation*}
where $\boldsymbol{a}_k = \Re\{\boldsymbol{w}_k\} $, $\boldsymbol{b}_k = \Im\{\boldsymbol{w}_k\}$. Introducing auxiliary matrix $\bold{C}_k  = \boldsymbol{a}_k \boldsymbol{a}_k^\mathrm{T} + \boldsymbol{b}_k \boldsymbol{b}_k^\mathrm{T}$ and  $\bold{D}_k =\boldsymbol{a}_k \boldsymbol{b}_k^\mathrm{T} - \boldsymbol{b}_k \boldsymbol{a}_k^\mathrm{T}$ for any $k \in \mathcal{K}$. Define angle related vectors where  $\boldsymbol{g}_k = \Re\{\bold a (\boldsymbol{d}, \phi_k)\}, \boldsymbol{q}_k =\Im\{\bold a (\boldsymbol{d}, \phi_k)\}, \boldsymbol{r} = \Re\{\bold a (\boldsymbol{d}, \phi)\}, \boldsymbol{u} = \Im\{\bold a (\boldsymbol{d}, \phi)\}$.

Furthermore, $\Pi_{\mathcal{C}_{\sf d}\cap ~\mathcal{C}_{\sf FA} \cap ~\mathcal{C}_{\sf SP}} (\bm{\kappa})$ can be computed by solving the following optimization problem.
\begin{equation}
	\min_{\bd} \|\boldsymbol{d} - \bm{\kappa}\|_2^2 \quad \text{s.t. } \boldsymbol{d} \in \mathcal{C}_{\sf d}\cap ~\mathcal{C}_{\sf FA} \cap ~\mathcal{C}_{\sf SP},
\end{equation}
where $\bm{\kappa}$ denotes the point to be projected. Since the constraint $\mathcal{C}_{\sf d}$ is nonconvex, we perform a quadratic taylor expansion on each cosine term in the power function at the current iteration point. To handle the nonconvex constraint $\bold a^\mathrm{H}(\boldsymbol{d}, \phi) \bold R_{w} \bold a (\boldsymbol{d}, \phi)$, we construct a concave lower bound using the second-order Taylor expansion of $\cos(\tilde{\phi}_{mn})$ around $\tilde{\boldsymbol{d}}$:
\begin{align}
	P(\boldsymbol{w}, \boldsymbol{d}, \phi) \geq \sum_{m,n} |R_{mn}| & \left[ \cos \tilde{\phi}_{mn} - \sin \tilde{\phi}_{mn} (\phi_{mn} - \tilde{\phi}_{mn}) \right. \nonumber \\
	& \qquad \left. - \frac{1}{2} (\phi_{mn} - \tilde{\phi}_{mn})^2 \right],
\end{align}
where $R_{mn}$ is the $m$th row and $n$th column element of matrix $\bold R_{w}$, for $n$, $m = 1,2,\dots,M$, and $\tilde{\phi}_{mn} = v(d_m -d_n + \angle R_{mn})$. 
To construct concave functions, linearization or quadratic approximation is required. Define $g(\boldsymbol{d}|\tilde{\boldsymbol{d}})$ as the lower bound of $\bold a^\mathrm{H}(\boldsymbol{d}, \phi) \bold R_{w} \bold a (\boldsymbol{d}, \phi)$
	\begin{align}
	\bold a^\mathrm{H}(\boldsymbol{d}, \phi) \bold R_{w} \bold a (\boldsymbol{d}, \phi) \geq g(\boldsymbol{d}|\tilde{\boldsymbol{d}}) \triangleq \boldsymbol{d}^\mathrm{T} \bold Q \boldsymbol{d} - 2\boldsymbol{d}^\mathrm{T} \boldsymbol{p} + c,
\end{align}
where $\boldsymbol{r} = \left[ \sum_{m=1}^M |R_{m1}|, \sum_{m=1}^M |R_{m2}|, \dots , \sum_{m=1}^M |R_{mM}|\right]$; $\bold Q=-v^2(\mathrm{diag} (\boldsymbol{r}-\bold R))$;  $[\bold R]_{mn} = |R_{mn}|$,
\begin{equation*}
	\begin{aligned}
	 \boldsymbol{p}[n] &= \sum_{m=1}^M |R_{mn}| \Bigl[ v  \sin\!\bigl(f(\tilde{d}_n,\tilde{d}_m)\bigr) - v^2 (\tilde{d}_n - \tilde{d}_m) \Bigr],\\
		c &= \sum_{m=1}^M \sum_{n=1}^M |R_{mn}|[\cos(f(\tilde{d}_n, \tilde{d}_m)) + v\sin(f(\tilde{d}_n, \tilde{d}_m)) \\
		& \qquad\qquad \times (\tilde{d}_n - \tilde{d}_m) - \frac{1}{2}v^2(\tilde{d}_n - \tilde{d}_m)^2],
	\end{aligned}	
\end{equation*}
 where $\tilde{\boldsymbol{d}} = \left[ \tilde{d}_1, \tilde{d}_2, \dots , \tilde{d}_M\right]^\mathrm{T}$; $f(\tilde{d}_n, \tilde{d}_m) = v(\tilde{d}_n, \tilde{d}_m) + \angle{R_{mn}}$  and $\tilde{\boldsymbol{d}}$ denotes any determined value for $\boldsymbol{d}$.  As a result, problem (21) is rewritten as follows:
\begin{align}
	\min_{\boldsymbol{d}} \|\boldsymbol{d} - \bm{\kappa}\|_2^2 \quad \text{s.t.} \; \boldsymbol{d} \in \mathcal{C}_p\cap ~\mathcal{C}_{\sf FA} \cap ~\mathcal{C}_{\sf SP},
\end{align}
where the feasible space $\mathcal{C}_p$ of $\boldsymbol{d}$ is restricted by $g(\boldsymbol{d}|\boldsymbol{d}^i)\geq P_d$.

The optimization problem (24) is a convex quadratic program with quadratic constraints (QCQP), solvable efficiently through readily available optimization tools such as CVX.

\subsection{Computational Complexity Analysis}

In this section, we analyze the computational complexity of the suggested BSUM algorithm. The total complexity of the BSUM algorithm is $\mathcal{O}(M^{3.5} + M^2(MK + M + K^2))$. In particular, the closed-form updates for $\mathbf{u}$, $\boldsymbol{\rho}$, $\boldsymbol{\xi}$, and $\boldsymbol{\eta}$ entail a complexity of $\mathcal{O}(M^2K^2)$, the beamformer optimization demands $\mathcal{O}(M^3 + M^2K)$, and the APV optimization encompasses $\mathcal{O}(M^{3.5} + M^2K^2)$.

\section{ SIMULATION RESULTS}
\begin{figure}[t]
	\centering
	\includegraphics[width=2.3 in]{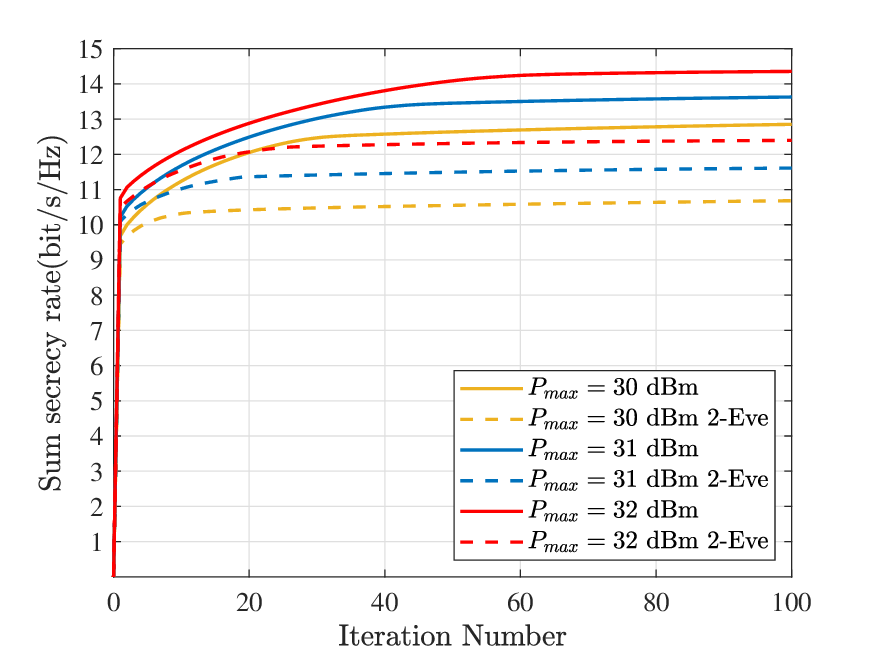} 
	\caption{Convergence of sum secrecy rate under different transmit power budgets with 2-Eve. $M= 8$, $K = 2$.}
	\vspace{-0.3cm}
	\label{fig:2}
\end{figure}

\begin{figure}[t]
	\centering 
	\vspace{-0.3cm}
	\subfloat[\scriptsize  $K$ = 2]{\includegraphics[width=0.50\columnwidth]{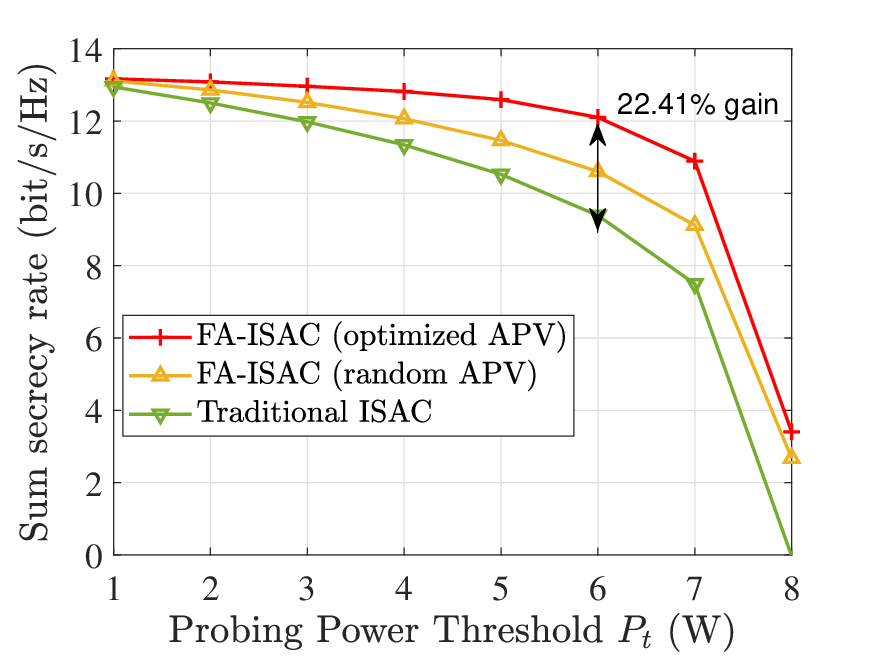}%
		\label{Pt_k2}}
	\hfil
	\subfloat[\scriptsize  $K$ = 8]{\includegraphics[width=0.50\columnwidth]{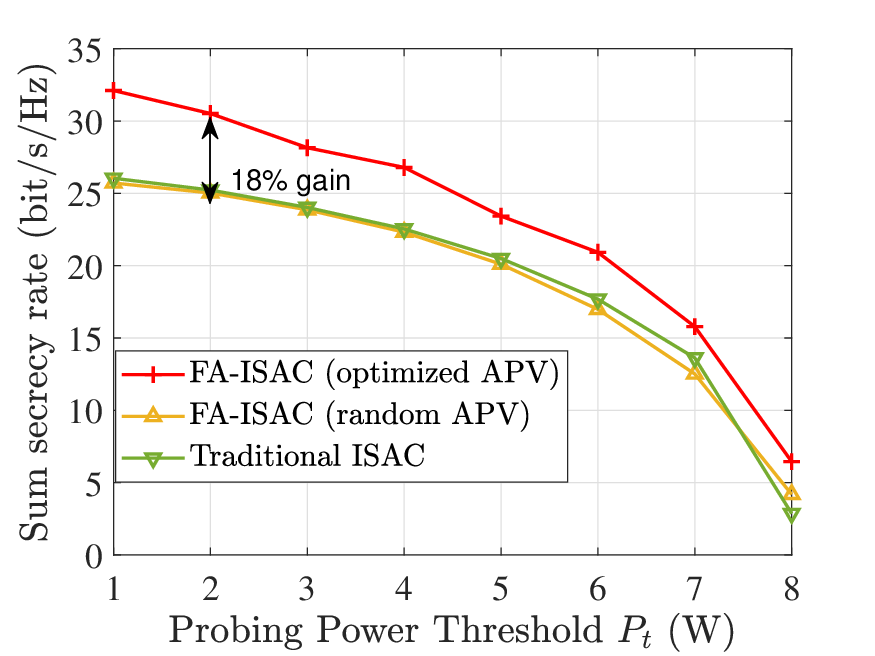}%
		\label{plot_Pt_k8}}
	\hfil
	\caption{Trade-off performance between sum secrecy rate and probing power $M= 8$, $P_{max} = 30$dBm.}
	\vspace{-0.2cm}
	\label{fig:4}
\end{figure}

\begin{figure}[t]
	\centering
	\vspace{-0.3cm}
	\subfloat[\scriptsize Sum secrecy rate] {\includegraphics[width=0.50\columnwidth]{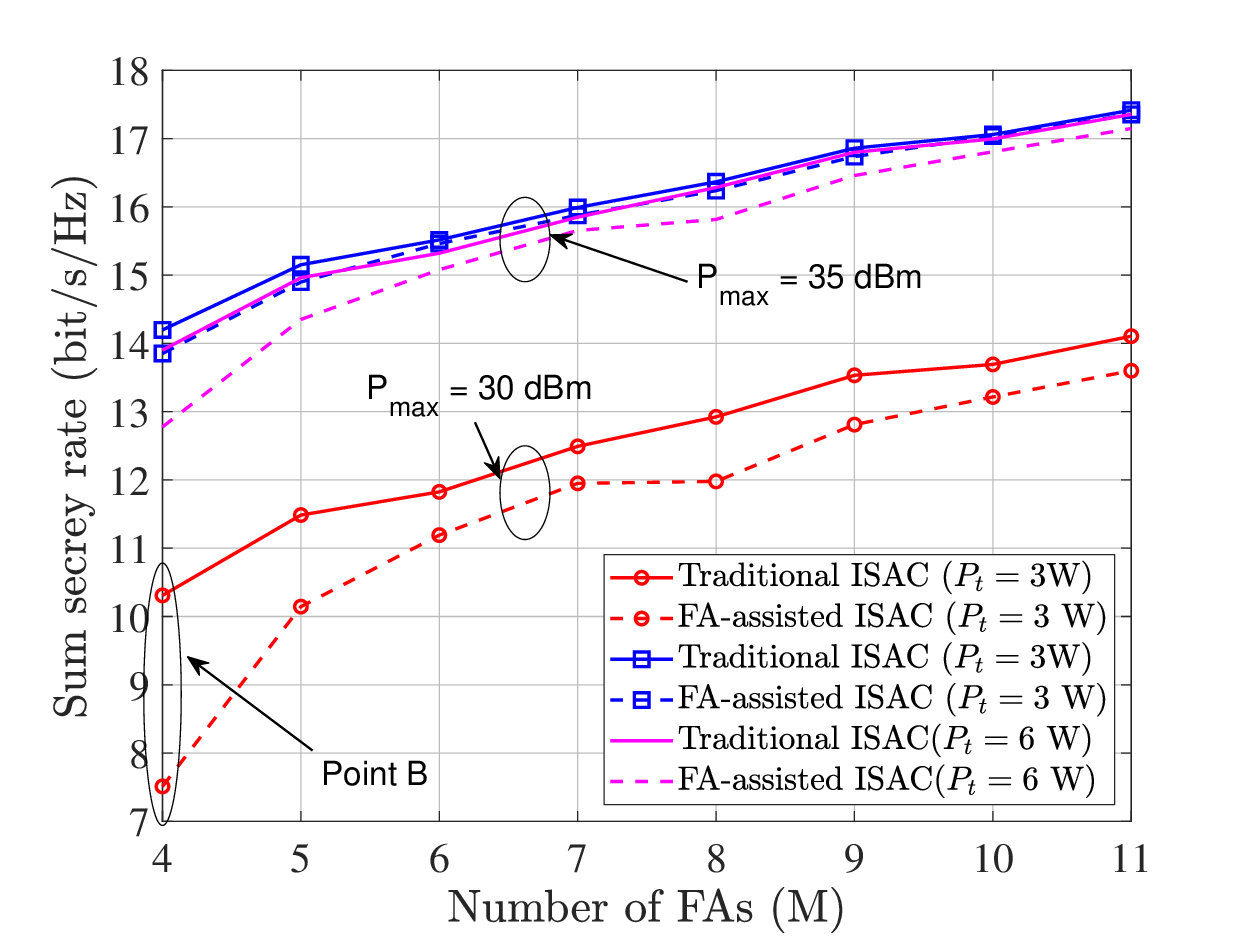}%
		\label{convergence_non_HWI}}
	\hfil
	\subfloat[\scriptsize Beampattern at Point B  ]{\includegraphics[width=0.50\columnwidth]{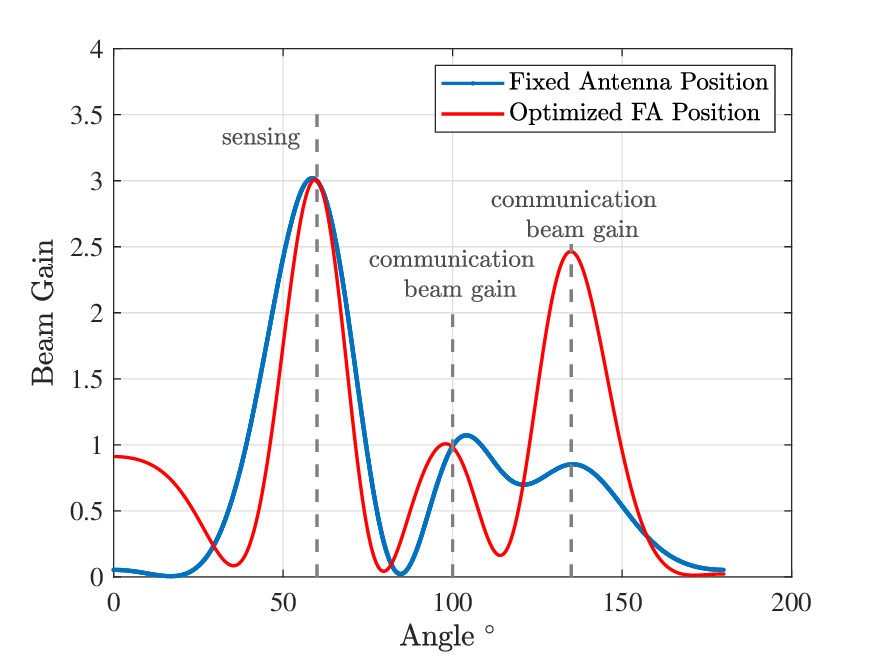}%
		\label{convergence_HWI}}
	\hfil
	\caption{Sum secrecy rate under different numbers of BS antennas. $K$=2.}
	\vspace{-0.2cm}
	\label{convergence}
\end{figure}

In this section, we conduct a numerical assessment of our proposed scheme for maximizing the sum secrecy rate in FA-aided ISAC systems. We configure the FA parameters as follows: $L_0 = \lambda/2$, $L = 10\lambda$, $\lambda = 0.01$ m, $\phi = 60^\circ$, $\sigma_k^2 = \sigma_e^2 = -80$ dBm, and $\delta_k = \delta_e = g_0 d_k^{-\alpha}$ for any $k \in \mathcal{K}$, where $g_0 = -40$ dB represents the fading at a reference distance of 1 meter and $\alpha = 2.8$ indicates the path-loss exponent. Users are located on a circle of radius 100 m centered at the BS, with $d_k = 100$m  for all $k \in \mathcal{K}$, and the range of $d_e$ is reasonably bounded. We analyze two scenarios: the 2-user and 8-user scenarios. In the 2-user case, the AODs of the FA array for the two users are $\phi_1 = 100^{\circ}$ and $\phi_2 = 130^{\circ}$, respectively. In the 8-user case, the AODs for the eight users are set to $\phi_1 = 10^{\circ}$, $\phi_2 = 30^{\circ}$, $\phi_3 = 80^{\circ}$, $\phi_4 = 90^{\circ}$, $\phi_5 = 120^{\circ}$, $\phi_6 = 130^{\circ}$, $\phi_7 = 150^{\circ}$, and $\phi_8 = 170^{\circ}$, respectively.

To evaluate the system robustness under more challenging conditions, Fig. 2 illustrates the convergence with two eavesdroppers. The algorithm converges within 200 iterations across power budgets, proving robustness. The secrecy rate increases with $P_{max}$. While extra eavesdroppers reduce the absolute rate due to worst-case constraints, FA-aided optimization effectively mitigates multi-directional leakage, maintaining a significant gain over conventional systems.

Fig. 3 shows the trade-off between the multiuser sum secrecy rate and probing power. It can be observed that in the $K=2$ scenarios with low sensing requirements, the FPA provides sufficient degrees of freedom (DoF) to support users, resulting in low enhancement of ISAC performance by the FA array. In contrast, in the $K=8$ scenarios, the FPA lacks adequate capacity to serve users, whereas the FA significantly enhances ISAC performance by offering greater spatial DoF. Under high sensing requirements, FAs demonstrate a substantial improvement in ISAC performance compared to the FPA in the $K=2$ scenarios, as the FPA struggles to effectively balance high probing power with multiuser communication demands. As shown in Fig. 3(b), the FA retains the capacity to balance both sensing, communication and secrecy rate, delivering significant performance improvements over the FPA.

Fig. 4(a) illustrates the impact of antenna number variation on the secrecy rate. When the number of antennas is small, the performance improvement from the FA becomes more pronounced, as the FA provides higher DoF to serve all users. Additionally, when $P_d = 3$ W, the performance gap between the FA and FPA arrays narrows as $P_{\max}$ increases, due to the expanded power budget providing greater power DoF. On the other hand, when $P_{\max} = 35$ dBm, this gap widens with increasing $P_d$, as the higher probing power makes the FPA array less capable of effectively serving users. Fig. 4(b) shows the beam pattern for point B in Fig. 4(a). FPA arrays cannot effectively guide communication beams in the desired direction while meeting sensing requirements. In contrast, by optimizing the FA position, the desired communication beam is achieved, resulting in a higher communication gain compared to the FPA.

\section{ CONCLUSION}
In this letter, we have investigated secure transmission in FA-aided ISAC systems, where the sensing target acts as a potential eavesdropper while exploiting spatial agility for channel reconfiguration. A multiuser sum secrecy rate maximization problem was formulated by jointly designing beamforming vectors and the APV, subject to probing power, position spacing, and total power constraints. To tackle the non-convexity, we used a BSUM-based algorithm, incorporating PDA for closed-form beamformer updates and EPG for APV optimization. Simulation results demonstrate the superiority of the proposed scheme, achieving over 20$\%$ secrecy rate gains versus FPA systems. Overall, our design can find applications in FA-aided ISAC systems with security considerations.

\begin{appendices}
	\section{}
Exchange the order of summation,
	\begin{align}
		\sum_{k=1}^{K} \sum_{i=1, i\neq k}^{K} [\boldsymbol{w}_i^\mathrm{H} \bold{E}_k(\boldsymbol{d}) \boldsymbol{w}_i] = \sum_{i=1}^{K} \sum_{k=1, k\neq i}^{K} [\boldsymbol{w}_i^\mathrm{H} \bold{E}_k(\boldsymbol{d}) \boldsymbol{w}_i].
	\end{align}
Taking the derivative of $\boldsymbol{w}_i$,
	\begin{align}
		\frac{\partial }{\partial \boldsymbol{w}^*_i} \sum_{k=1, k\neq i}^{K} [\boldsymbol{w}_i^\mathrm{H} \bold{E}_k(\boldsymbol{d}) \boldsymbol{w}_i] = \sum_{k=1, k\neq i}^{K} \bold{E}_k\boldsymbol{w}_i,
	\end{align}
we define $\sum_{k=1, k\neq i}^{K} \bold{E}_d = \bold{E}(\boldsymbol{d}) - \bold{E}_i(\boldsymbol{d})$,
and provide the following expression as
\begin{align}
	\frac{\partial }{\partial \boldsymbol{w}^*_i} \sum_{k=1, k\neq i}^{K}  [\boldsymbol{w}_i^\mathrm{H} \bold{E}_k(\boldsymbol{d}) \boldsymbol{w}_i] &=  (\bold{E}(\boldsymbol{d}) - \bold{E}_i(\boldsymbol{d}))\boldsymbol{w}_i.
\end{align}

\end{appendices}

%


\vfill

\end{document}